\documentclass{raa}
\usepackage{graphicx,times}             
\usepackage{natbib}
\usepackage{multicol}
\usepackage{amssymb,amsmath}
\usepackage{color}

\newcommand{\erg}{${\rm erg \ s^{-1}}$ }

\def\ltsima{$\; \buildrel < \over \sim \;$}
\def\simlt{\lower.5ex\hbox{\ltsima}}
\def\gtsima{$\; \buildrel > \over \sim \;$}
\def\simgt{\lower.5ex\hbox{\gtsima}}
\newcommand{\msun}{{\rm\,M$_\odot$}}

\newcommand{\srcs}{{\rm\,AT2018cow}}
\newcommand{\src}{{\rm\,AT2018cow }}
\newcommand{\xmm}{{XMM-Newton} }
\newcommand{\xmmp}{{XMM-Newton}}
\newcommand{\swift}{{Swift} }
 \newcommand{\swiftp}{{Swift}}
\usepackage{threeparttable}
\usepackage{bibentry}

\begin{document}

\title{
A possible 250-second X-ray quasi-periodicity in the fast blue optical transient AT2018cow
}

\author{Wenjie Zhang\inst{1}
\and Xinwen Shu\inst{1}
\and Jin-Hong Chen\inst{2}
\and Luming~Sun\inst{1}
\and Rong-Feng Shen\inst{2}
\and Lian Tao\inst{3}
\and Chun Chen\inst{2}
\and Ning~Jiang\inst{4}
\and Liming Dou\inst{5}
\and Ying Qin\inst{1}
\and Xue-Guang Zhang\inst{6}
\and Liang Zhang\inst{3}
\and Jinlu Qu\inst{3}
\and Tinggui~Wang\inst{4}
}
\institute{Department of Physics, Anhui Normal University, Wuhu, Anhui, 241002, China; {\it xwshu@ahnu.edu.cn} \\
\and School of Physics and Astronomy, Sun Yat-Sen University, Zhuhai 519082, China;  {\it shenrf3@mail.sysu.edu.cn} \\
\and Key Laboratory of Particle Astrophysics, Institute of High Energy Physics, Chinese Academy of Science, Beijing 100049, China \\
\and CAS Key Laboratory for Researches in Galaxies and Cosmology, Department of Astronomy,  University of Science and Technology of China, Hefei, Anhui 230026, China; {\it twang@ustc.edu.cn}  \\
\and Department of Astronomy, Guangzhou University, Guangzhou 510006, China \\
\and School of Physics and Technology, Nanjing Normal University, Nanjing 210023, China}

\abstract{  The fast blue optical transients (FBOTs) are a new population of extragalactic transients of unclear physical origin. 
A variety of mechanisms have been proposed including failed supernova explosion, shock interaction 
with a dense medium, young magnetar, accretion onto a compact object, and stellar tidal disruption event, 
but none is conclusive. 
 Here we report the discovery of a possible X-ray quasi-periodicity signal with a period of $\sim$250 second (at a significance level of 99.76\%)
in the brightest FBOT AT2018cow through the analysis of XMM-Newton/PN data. 
The signal is independently detected at the same frequency in the average power density spectrum from data taken from the \swift telescope, with observations covering from 6 to 37 days after the optical discovery, though the significance level is lower (94.26\%).
This suggests that the QPO frequency may be stable over at least 1.1×10$^{4}$ cycles.
Assuming the $\sim$250 second QPO to be a scaled-down analogue of that typically seen in stellar mass black holes, a black hole mass of $\sim10^{3}-10^{5}$ solar masses could be inferred. 
 The overall X-ray luminosity evolution could be modeled with the stellar tidal disruption by a black hole of $\sim10^4$ solar masses, 
 providing a viable mechanism to produce AT2018cow.    
 Our findings suggest that other bright FBOTs 
 may also harbor intermediate-mass black holes. 
}


   \maketitle

\section{INTRODUCTION}

Recent optical time-domain surveys have discovered a new population of fast-rising blue optical transients (FBOTs). 
These objects are characterized by rapid rise to their peak brightness within $\simlt$10 d and 
blue colors ($g-r<-0.2$) near the peak, followed by fading away of emission within $\simlt$100 d      (\citealt{Drout2014, Pursiainen2018, Ho2021}). 
\src is one of the most extreme FBOTs, hosted in the dwarf spiral galaxy CGCG 137-068 at a luminosity distance 66 Mpc ($z=0.0141$, \citealt{Prentice2018}). 
After its discovery by ATLAS survey on 2018 June 16,  \src received prompt and extensive multi-wavelength observations 
spanning from radio to $\gamma$-rays (\citealt{Sandoval2018, Ho2019, Margutti2019, Perley2019, Roychowdhury2019,  Kuin2019, Mohan2020, Sun2022}), 
confirming it to be the brightest FBOT known so far with a peak bolometric luminosity of $4\times10^{44}$ \erg, 
that locates outside the nucleus of the host galaxy with a positional offset by 1.7 kpc. 
At the explosion site of \srcs, a young star population in dense gas environment was inferred, 
indicating active star-formation (\citealt{Morokuma2019, Lyman2020}). 
Despite its physical origin remains controversial, the multi-wavelength analysis, especially those in the X-ray bands, 
indicates that \src can be powered by a compact object (\citealt{Margutti2019}), either newly formed stellar-mass black hole or 
neutron star in a supernova, or tidal disruption of a star by an intermediate-mass black hole (IMBH, a few 
$10^{4}-10^{5}$ solar masses) (\citealt{Perley2019, Kuin2019}).  
The detection of 4.4ms quasi-periodic oscillation (QPO) from NICER observations 
argues further for the presence of a compact object in \src (\citealt{Pasham2021}), 
but the study is limited to only the soft X-ray band (0.25--2.5 keV) because of 
 the high background contamination beyond $\sim$2.5 keV. 
It is still not conclusive whether the compact object is a neutron star or a black hole shining \srcs. 

In this paper, we report the detection of a possible X-ray QPO signal at $4.07 \pm 0.39$ mHz 
from 
XMM-Newton and Swift observations of AT2018cow when its X-ray emission is in the relatively bright phase ($L_{\rm 0.3-10 keV}\simgt10^{42}$\erg).
We also propose a model with the tidal disruption by an IMBH with mass of $\sim10^{4}$\msun to explain the unusual X-ray light curve of AT2018cow. 
In Section 2, we describe the observations and data reduction. 
Section 3 presents the detailed timing analysis and results. 
Possible origins of the QPO at $\sim$4.1 mHz and its implications on the 
central engine of \src are discussed in Section 4. 
\section{Observations and DATA REDUCTION} 



\begin{figure*}[htb]
\centering
\includegraphics[width=\linewidth]{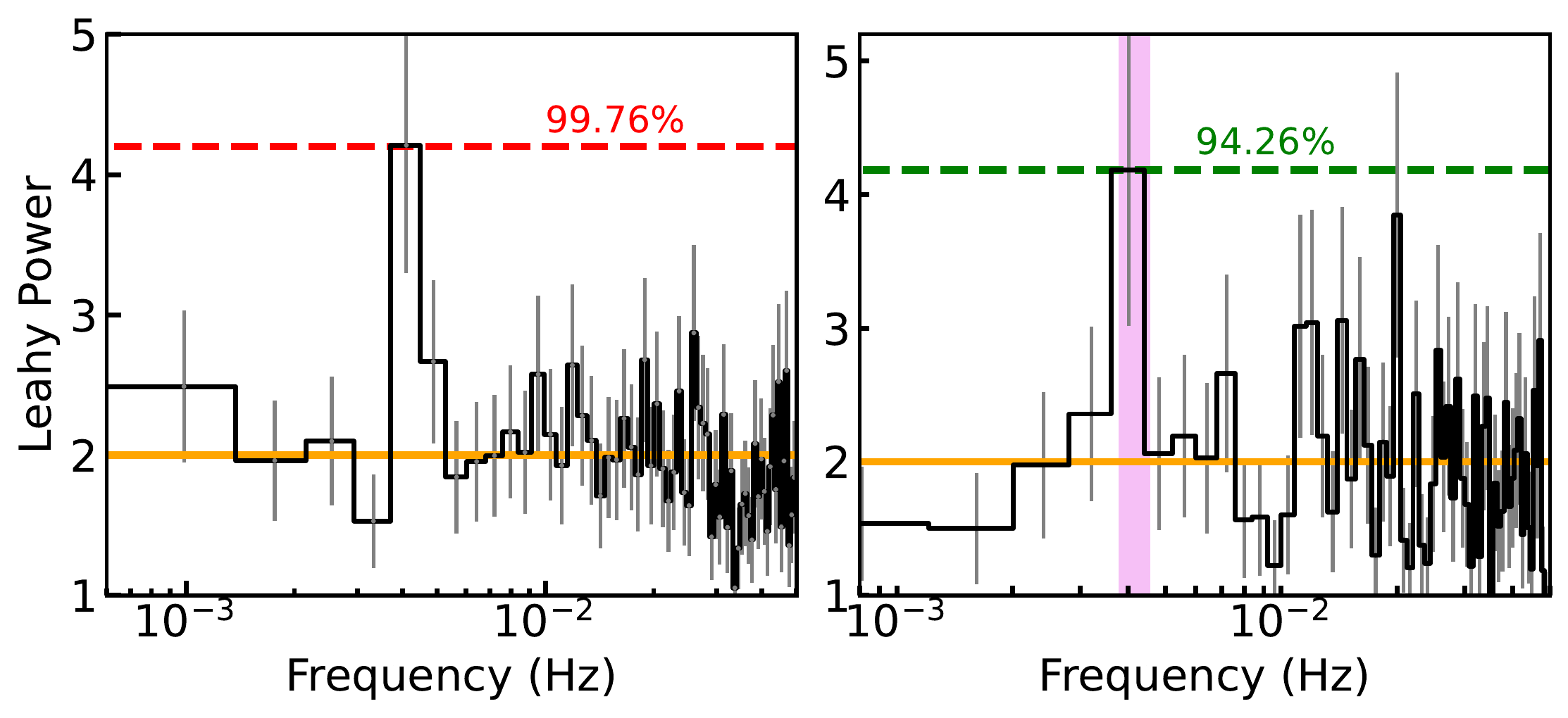}
\caption{XMM-Newton and Swift power density spectrum for AT2018cow. 
{\it Left}: XMM-Newton Leahy-normalized PDS of AT2018cow in the 0.3-10 keV. The PDS was binned to have a frequency resolution of 0.78 mHz. A strong peak appears at $4.07\pm0.39$ mHz 
($\approx $246-second), indicating the presence of a QPO component. 
The QPO frequency is defined as the centroid value of the peak frequency bin, and the error is the half of the bin width.   
The red dashed line represents 
the 99.76\% confidence level derived from Monte Carlo simulations. 
The orange horizontal line shows the white noise (the value is 2 in Leahy-normalized power spectrum). 
{\it Right}: Averaged Swift PDS with a frequency resolution of 0.8 mHz. The green dashed line represents 94.26\% confidence level. The purple strip shows the location of QPO in XMM-Newton. 
The error bars represent 1$\sigma$ uncertainties.  \label{fig:f1}}
\end{figure*}

AT2018cow has been observed by XMM-Newton on three epochs, about 37 days, 82 days, and 218 days after the optical discovery. 
However, only the data from the first \xmm observation have enough photons for detailed timing analysis. We used principally 
the data from the EPIC PN camera, which have a much higher sensitivity. We processed data using Science Analysis Software version 17.0.0, with the latest calibration files.  
 We extracted the event files from a circular region with a 35$^{\prime \prime}$ radius centered on the source position from 
 optical observations (R.A.: 16$^{\rm h}$16$^{\rm m}$00$^{s}$.22, Dec.: +22$^{\circ}$16$^{\prime}$04$^{\prime \prime}$.8, \citealt{Prentice2018}) {and the background was extracted from four source-free circular regions with 35$^{\prime \prime}$ radius near the source position.}
Both the source spectrum and light curve in the energy range of 0.3-10 keV were selected, and only the good events PATTERN$\leq 4$  were used in generating light curve. Then we used {\tt epiclccorr} tool to correct the light curve for instrumental effects. 
{By examining the background light curve, we found that background flares were present at the beginning 
of the observation. We ignored the first $\sim$3 ks to remove the time interval that may be affected by 
the high particle background, resulting in a net exposure of $\sim$27 ks. 
We assessed the extent of photon pile-up using the SAS task {\tt epatplot}, and found that such 
an effect is negligible.  
 
AT2018cow has also been intensively observed by Swift-XRT covering a period of $t=3-70$ days since the optical discovery\footnote{We do not use the six Swift-XRT observations 
at later times ($t=120-1380$ days), since the source has decayed to a flux level not detectable with \swiftp. }. 
Totally there are 95 individual Swift-XRT observations in which \src is detected, which provide critical constraints on the long-term evolution 
of its X-ray emission. } For Swift-XRT data, we used the task {\tt xrtpipline} to generate the event files, and then used {\tt XSELECT} in HEASoft\ 
to extract the source spectrum in a circular region with a radius of 40$^{\prime \prime}$, and the background spectrum in an annulus with an inner radius 
of 60$^{\prime \prime}$ and outer radius of 110$^{\prime \prime}$, respectively. All spectra were extracted in the energy of 0.3-10 keV.  The light curves were corrected for bad time intervals using the task {\tt xrtlccorr} and the background was subtracted by using standard FTool {\tt lcmath}.
Owing to a few counts in individual observations, we cannot derive source's flux through spectral fittings. 
We used {\tt WebPIMMS} tool to convert the Swift-XRT 
count rates to flux by assuming an absorbed power-law model (\citealt{Margutti2019}) with photon index $ \Gamma=1.6$ 
modified by a Galactic HI column density of $N_{\rm H}=0.05 \times 10^{22}$ cm$^{-2}$.  
 
 \begin{figure*}[htb]
\centering
\includegraphics[width=\linewidth]{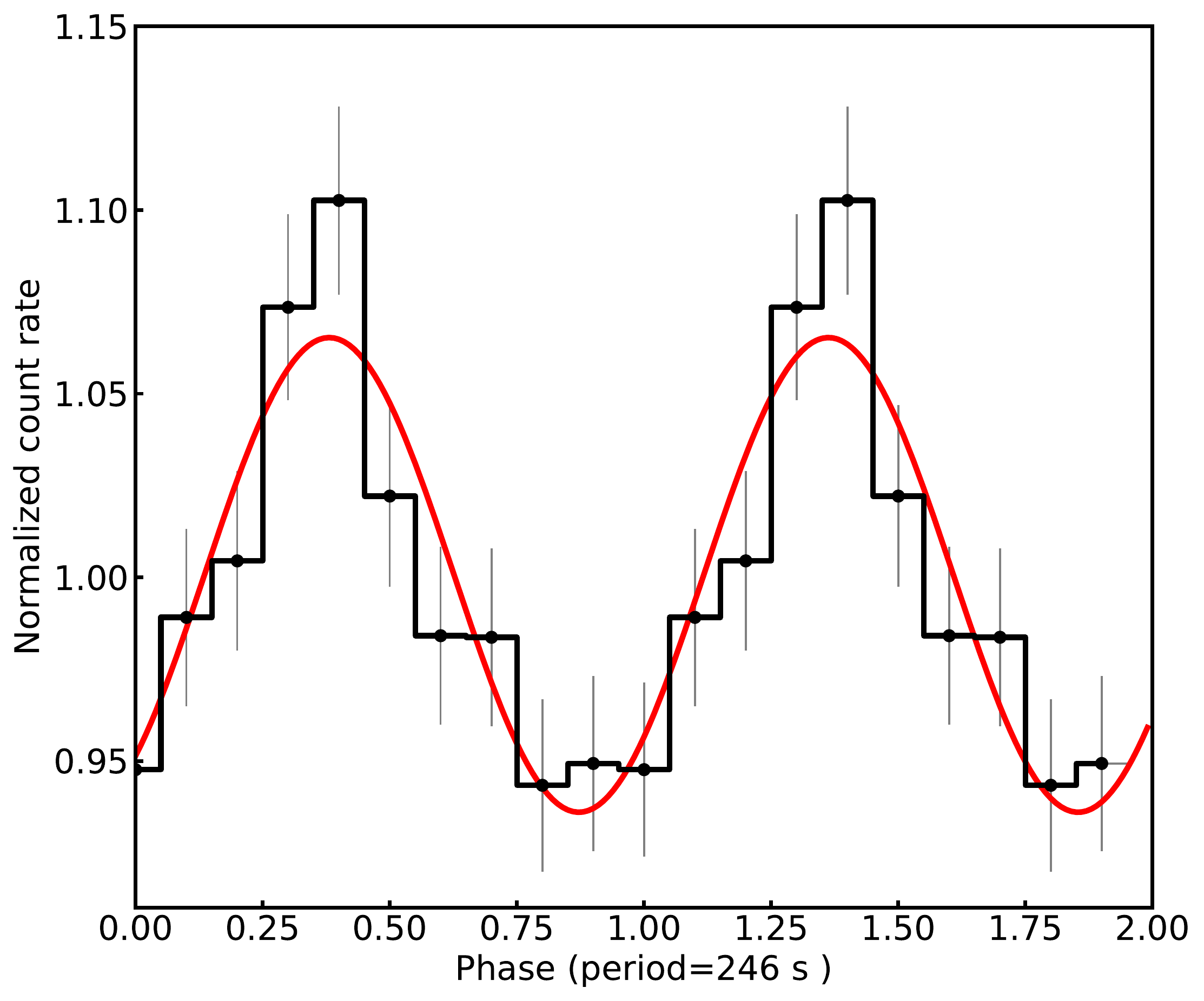}
\caption{XMM-Newton folded light curve in the 0.3-10 keV with a period of 246 second, and the best-fit sinusoid is shown in red solid line. The error bars represent 1$\sigma$ uncertainties.}
\end{figure*}

\section{Light curve analysis and results} \label{sec:style}


The PN light curve was rebinned to have a time binsize of 10 s. 
Given the mean count rate of 0.62 cts/s, this ensures that there are enough photons in each time bin for meaningful 
timing analysis.  
We then Fourier transformed the light curve into power density spectrum (PDS) which was normalized by Leahy method  (\citealt{Leahy1983}) to have a mean Poisson noise level of 2. 
The Leahy-normalized PDS was then rebinned by a factor of 21, which gives a frequency resolution of $\sim$0.8 mHz. 
Figure 1 (left panel) shows the resulting PDS, which reveals an apparent QPO component at a frequency of $4.07 \pm 0.39$ mHz, 
corresponding to a period of 246 s.  {We repeated the above analysis by changing the size and position of both source and  
background extraction areas, and found that the PDS are basically unchanged. In addition, we also checked the PDS of the 
background light curve and found that there is no QPO signal, so we excluded the possibility that the QPO signal at 4.07 mHz comes 
from background fluctuations. 
 Furthermore, we} checked that the power value ($\xi_{obs}=4.2$) at the frequency of $\sim$4.1 mHz does not change if the PDS was constructed 
from the light curve with smaller time binsizes ($<$10 s). 
The QPO signal appears to present over the entire energy band of our interest (0.3--10 keV), as its power increases 
gradually if more photons at higher energies are included (Figure A1 in Appendix). 
However, if focusing on only the hard X-ray bands at $>$2 keV, the QPO feature disappears, 
suggesting that the signal is dominated by photons at soft bands.

In order to confirm the QPO detected in the XMM-Newton data, we also performed PDS analysis for data taken from Swift-XRT because its effective energy band is similar to that of XMM-Newton. Due to the individual Swift-XRT observations are short and most have an exposure less than 1 ks, 
we selected the observations with continuously effective exposure time of $>$1.25 ks and count rates $>0.1$ cts s$^{-1}$ 
to ensure enough counts for timing analysis. 
This results in a total of 13 useful light curves (Table B1 in Appendix). 
 We further restricted all individual light curve segments to have the same exposure of 1.25 ks for stacking the PDS. 
 The reason to choose the length of 1.25 ks is to ensure that the frequency width of PDS (1/1250s) 
 is close to that of PN (21/27000s). 
 Since most of usable Swift-XRT observations have a length of 1.3--2 ks, 
 we uniformly chose the initial 1.25 ks in each light curve. 
 This will help to avoid the effect of arbitrarily selected light curve segments that may cause false signals in the PDS. 
Note that we split one light curve that has enough exposure length (2.8 ks) into two 1.25 ks segments. 
The light curve segments were rebinned to have a time binsize of 10 s, resulting in 
at least 1 count per bin, which is sufficient for a reliable timing analysis. 
By stacking the PDS from totally 13 individual light curve segments each with 1.25 ks, 
we extracted an average PDS for the Swift-XRT data. 
The Leahy-normalized PDS is shown in Figure 1 (right panel) where an obvious feature 
is observed at 4.0$\pm$0.4 mHz, which is consistent with the QPO frequency obtained with the \xmm data within errors. 
{Note that the Leahy-normalized PDS was produced without weighting by flux for the Swift-XRT data, as it 
is not clear whether there is a strong dependence of QPO strength on the X-ray flux.
}

We also used the light curve analysis tool {\tt efold} in HEASoft to fold the PN light curve with the period of  246 s, 
which is shown in Figure 2.
The fractional rms amplitude calculated by the folded light curve is $6.0\pm1.3$\%.  
The light curves from individual \swift observations do not have enough signal-to-noise ratios to perform the folded analysis. 
Therefore, we did not estimate the QPO rms amplitude for the Swift data.

 To estimate the statistical significance of the QPO detected at $\sim$4.1 mHz which depends on the underlying distribution of the noise powers, 
 an important question is to test whether the noise at other frequencies (except for the frequency where the QPO is located) in the PDS is consistent with white noise which is relatively flat, or red noise whose strength depends on frequency. A direct way to distinguish the type of noise is to examine whether the noise powers are $\chi^{2}$ distributed as expected for white noise. 
We removed the three frequency bins centered at 4.1 mHz, and generated a cumulative distribution function of the powers at all other frequencies. 
We then compared it with a $\chi^{2}$ distribution (with $2\times21$ degrees of freedom, d.o.f) scaled by a factor of 1/21, where the number 21 is the factor by which the PDS was rebinned,  
as shown in the Figure C1 in Appendix. 
In addition, we also analyzed the probability density distribution of these powers and compared it with the expected values from a $\chi^{2}$ distribution. 
Both tests show consistent results that the powers of the noise can be described with $\chi^{2}$ distribution, i.e., 
dominated by white noise. 
For the \swift PDS, we also tested that the noise powers are $\chi^{2}$ distributed (Figure C1 in Appendix). 
Therefore, for the white noise distribution of the underlying continuum in the PDS, 
the significance of the QPO at $\sim$4.1 mHz can be analytically estimated using the $\chi^{2}$ distribution. 
For the \xmm data, we found that the probability to observe a power value larger than measured ($\xi_{obs}=$4.2), is 
P$(>$$21\xi_{obs}; \rm d.o.f)=3.9\times10^{-5}$$=p_{\rm 1}$. 
By taking into account the search for all frequency bins over the range of interest ($N=64$), 
the global significance is 1-$p_{\rm 1}\times$N=99.75\%. 
Similarly, for the Swift PDS, the probability to observe a power value larger than measured ($\xi_{obs}=$4.18), is 
P$(>$$13\xi_{obs}; \rm d.o.f)=9.26\times10^{-4}$$=p_{\rm 2}$.  
The global confidence level for N = 62 frequency trials is 1-$p_{\rm 2}\times$N=94.26\%.

\begin{figure*}[t]
\centering
\includegraphics[width=\linewidth]{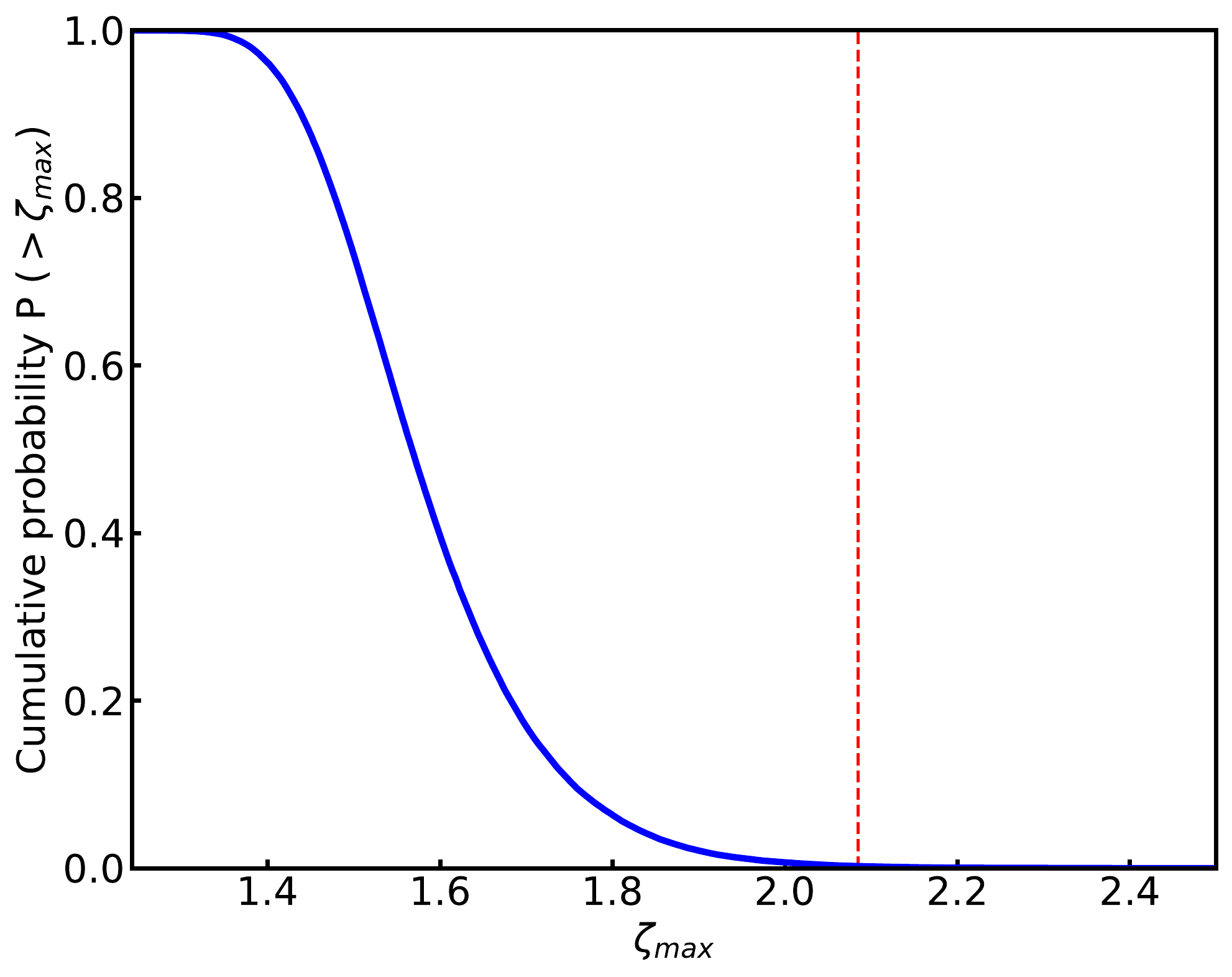}
\caption{Cumulative distribution of the maximum noise power $P(>\xi_{max})$ from Monte Carlo simulations. 
The red dashed line marks the $\xi_{max,obs}$ measured from the \xmm data. 
By comparing  $\xi_{max,obs}$ with the cumulative distribution of the noise power, we 
derived a global statistical significance of 99.76\% for the QPO detected in the \xmm data.  
 All the power values were divided by the average value at the corresponding frequency bin (Section 3 for details). 
}
\end{figure*}

The above assessment of noise distribution in the PDS is only qualitative. 
It is still possible that a weak or unknown red noise component is present in the data. 
To estimate the QPO significance by properly accounting for the underlying noise, we then employed a Monte Carlo approach, 
by generating a series of light curves following a given power spectral distribution (\citealt{Uttley2002}). 
We used the powerlaw plus constant model, $ P(f) = Nf^{-\alpha} + C $, to fit the unnormalized and unbinned PDS of XMM-Newton, where N is the normalization factor,  $ \alpha $ is the powerlaw index, and C is a constant indicating the Poisson noise level. The powerlaw component was used to account for possible effect of weak red noise on the simulation results. 
Using the maximum likelihood estimation method (\citealt{Vaughan2010}), we first obtained the best fitting parameters, where 
N = -0.71, $\alpha$ = -1.1$\times10^{-5}$ and C = 5.04.
Then we used the best-fit PDS model to generate a series of simulated light curves (\citealt{Timmer1995}), which were resampled to have the same duration, mean count rate, and variance as the real light curve. To eliminate the possible effect of red noise leakage at the edge of light curve, we choose a length of $5\times 27000$ s which is five times longer than the real data.  Then we intercepted the middle 27 ks as the final light curve for simulations.  
We repeated this process and obtained 100,000 simulated light curves. The same power spectral analysis on these simulated light curves was performed as we did on the real data, producing 100,000 simulated PDS.

With the 100,000 simulated PDS, we computed the significance level of a QPO peak by scanning all frequency bins below 0.05 Hz, the highest 
frequency in the PDS. We averaged the power of the 100,000 PDS at each frequency bin and obtained the $ 	\left \langle P_{f_{i}} \right \rangle$ where $i$ is $i^{\rm th}$ number of bins ($i=1, \dots, n$). 
For each simulated PDS, we divided the power value at each frequency by the average value $ 	\left \langle P_{f_{i}} \right \rangle$ at the corresponding frequency, and recorded the largest value as $\xi_{max}$, yielding 100,000 $\xi_{max}$ at a given frequency. Averaging the power distribution has the advantage of eliminating the possible effect of red noise (\citealt{Pasham2019}). 
Using the 100,000 values of $\xi_{max}$, we calculated the cumulative 
distribution of probability to exceed a given $\xi_{max}$ (Figure 3). 
By comparing with the observed value of $\xi_{max,obs}$, we 
obtained the global confidence level of 99.76\% for the QPO at $\sim$4.1 mHz.
Using the same Monte Carlo approach, we obtained that the statistical significance of the QPO signal in the \swift PDS is {94.38\%}. 
The results from Monte Carlo simulations are in good agreement with that obtained with the $\chi^{2}$ analysis of white noise distribution, 
indicating that the red noise component (if present) is negligible. 
 Considering that the QPO was detected at the same frequency in two independent detectors, 
 its combined confidence level is 99.986\%, 
 making the signal statistically significant. 
Note that we did not detect the QPO signal in either individual or combined EPIC MOS data, possibly due to the 
relatively low count rates in the individual MOS light curves, which are a factor of $\sim$3.5 
less than that in PN. In addition, the fractional rms amplitude is only 6\%, making 
that the QPO signal (if presents) could be more affected by noise fluctuations in the MOS data. 
In order to test this possibility, we performed detailed simulations in Appendix D, and found 
that the chance of the same QPO signal detected at a level of {99.76\%} in the MOS data is indeed low ($<$0.1\%).  
{We also performed the PDS analysis by combining PN and MOS light curves, and found that 
the power value at 4.1 mHz comes down, 
probably due to the fact that QPO signal is not detectable in the MOS data. 
In this case, the global confidence level for the QPO in the combined light curve is reduced to 92.81\% (Fig. D2).}

\section{Discussion and conclusions} 
Having established that the QPO is statistically significant, 
we now place its constraint on the origin of the X-ray emission and 
the central engine of \srcs. 
On a basis of the soft X-ray QPO identified at 224 Hz, 
a compact object for the origin of X-ray emission was proposed (\citealt{Pasham2021}). 
The discovery of $\sim$4.1 mHz QPO 
provides further evidence for the presence of a compact object, which could be either a neutron star, or an accreting black hole. 
mHz QPOs have been detected in a few neutron-star low-mass X-ray binaries (\citealt{Revnivtsev2001, Tse2021}), 
which are generally explained as thermonuclear burning on the neutron star surface. 
As a result, these QPOs are transient behavior occurring only in a narrow range of X-ray luminosity during outbursts. 
This contradicts the constant frequency of the QPO at $\sim$4.1 mHz in AT2018cow, 
as the stacking analysis of \swift data suggests that the QPO frequency may be stable over a period of 30 days, 
during which the luminosity declined by a factor of $>$5. 
The QPO's stability may also challenge other models involving a neutron star (\citealt{Pasham2021}), 
such as spin-down of an isolated millisecond magnetar or magnetar-accretion driven outburst. 

If the compact object was a stellar-mass black hole, the 4.1 mHz QPO is unlike the typical low-frequency 
QPOs (LFQPOs), as the frequency is lower than most of the type-A ($\sim$8 Hz), type-B ($\sim$5-6 Hz), 
and type-C ($\sim$0.1-15 Hz) QPOs (\citealt{Casella2005, Remillard2006, Ingram2019}). The QPO's frequency is also far less than 
a few 10-100 Hz for the high-frequency QPOs (HFQPOs) (\citealt{Morgan1997, Belloni2012}). Although several accreting stellar-mass black holes have shown 
QPOs with frequency in the mHz range (\citealt{Cheng2019}), most of these sources are believed to have 
high orbital inclinations. This is inconsistent with the weak or no X-ray absorption in observations on 
time-scales of hours to days (\citealt{Margutti2019}). 
Furthermore, almost all these sources have shown the red noise PDS at below 1 mHz. 
These properties distinguish them from the 4.1 mHz QPO found in \srcs, which appears with weak 
or absent red noise. 
Two black hole sources, GRS 1915+105 and IGRJ17091--3624, show mHz QPOs in the so-called ``heartbeat''
state (\citealt{Belloni2000, Altamirano2011}).  
However, these QPOs are characterized by high harmonic content and accompanied with flat-top noise in the PDS 
where a low and high frequency break are evident. 
On the other hand, if isotropically emitted, the peak luminosity of \src is $\simgt10^{4}$ times the Eddington limit for a stellar-mass 
black hole ($M_{\rm BH}\sim10$\msun), challenging standard models of black hole accretion.  
Therefore, if the mHz QPO of \src was due to a process similar to that operating in stellar-mass black holes, 
the underlying physical mechanism would be extreme and have never been seen before. 

\begin{figure*}[ht]
\centering
\includegraphics[width=\linewidth]{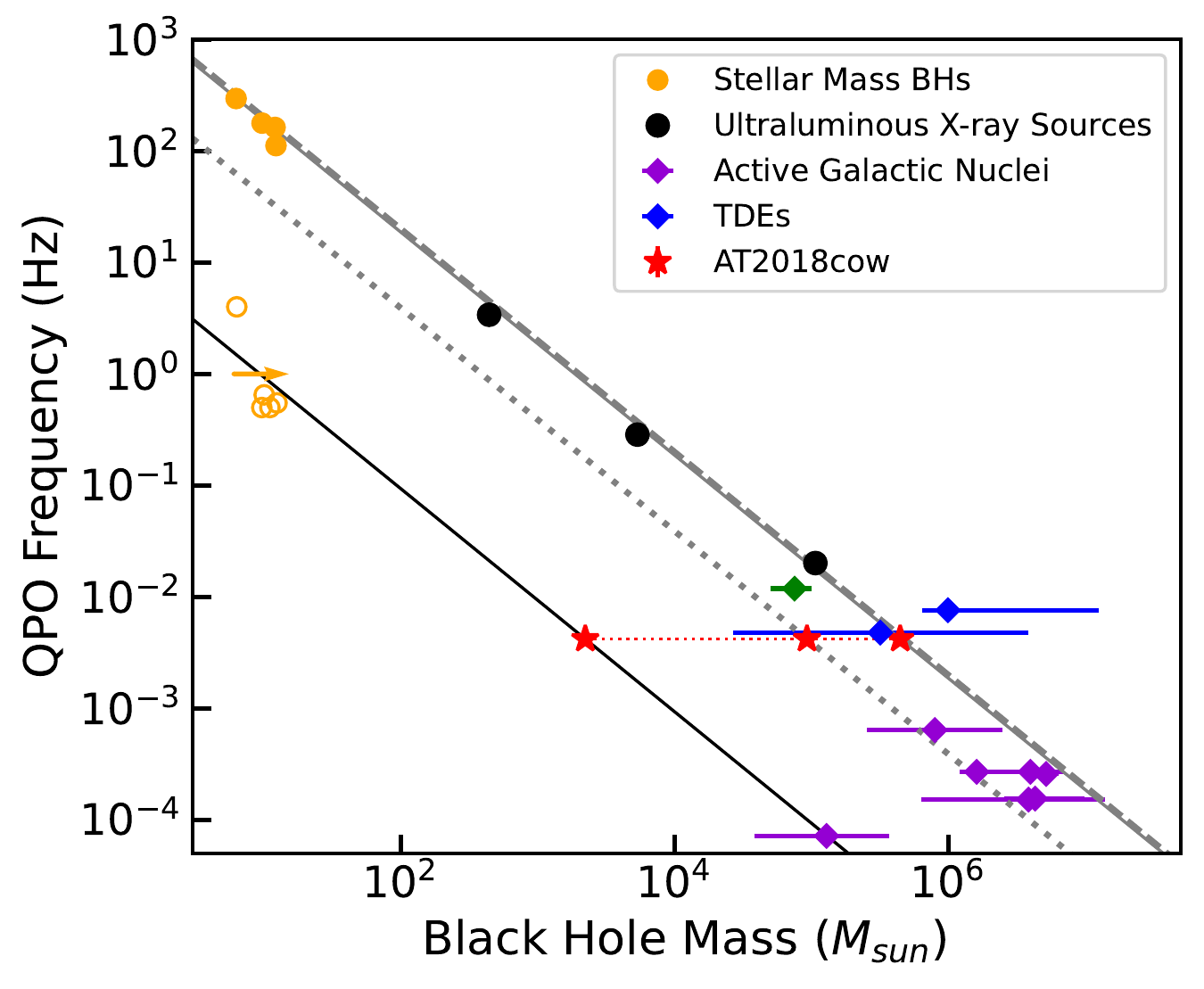}
\caption{Relation between QPO frequency and BH mass.
The gray solid line is the extrapolation of the relation between QPO frequency and BH mass for stellar mass black holes derived in (\citealt{Remillard2006}), 
assuming that the QPO corresponds to the 2$\times f_{0}$ for the 3:2 harmonic peaks. 
The dashed and dotted line represent the relation derived from the model of 3:2 resonance  (\citealt{Aschenbach2004}) with the
spin parameter $a=0$ and $a=0.998$, respectively.
The HFQPOs are shown in filled orange circles (\citealt{Zhou2015, Strohmayer2009}), while LFQPOs are represented with open orange circles. 
The frequencies of LFQPOs are derived using the relation between frequency and spectral index (\citealt{Shaposhnikov2009}), with the 0.3--10 keV 
spectral index ranges from 1.6 to 1.7 as observed in \srcs. Only the stellar-mass black holes with dynamic mass constraints are displayed. 
The data for ultraluminous X-ray sources harboring IMBHs, TDEs (ASASSN-14li and Sw J1644+57) and AGNs with supermassive BHs are taken 
from (\citealt{Pasham2015, Reis2012, Pasham2019, Song2020}). 
The gray solid line is scaled downward by a factor of 200 to approximately match the distribution of LFQPOs. 
The \srcs' BH mass range constrained by the 4.1 mHz QPO is shown by red stars. 
The errors on BH mass are reported to take into account the statistically errors and typical errors of $\sim$0.5 dex of the scaling relation 
between the BH mass and host properties. 
}
\end{figure*}

The observed frequency, if it is related to the Keplerian period of the innermost circular stable orbit, 
would correspond to a black hole mass of between $5\times10^{5}$\msun~and $6\times10^{6}$\msun, 
for a non-rotating and maximally rotating black hole, respectively. 
Alternatively, the 4.1 mHz QPO could represent a scaled-down analogue of the typical LFQPOs or HFQPOs of stellar-mass black holes, 
but occurring at a lower frequency if a more massive BH or IMBH is at work in \srcs.  
The frequencies of HFQPOs appear to be stable and scale inversely with mass, $f_{\rm 0}=931 (M/$\msun)$^{-1}$ Hz (\citealt{Remillard2006}). 
Assuming the 4.1 mHz corresponds to the stronger $2\times f_{\rm 0}$ frequency, the scaling relation yields a black hole mass 
of 4.4$\times10^5$\msun~in \src (Figure 4).  
As we mentioned above, different types of LFQPOs are observed in the frequency range of $\sim0.1-20$ Hz. 
The QPO frequency can be variable and correlated with the energy spectral parameters, i.e., the flux and shape of 
power-law spectral component (\citealt{Remillard2006}). During the first 37 days, the 0.3--10 keV spectrum of \src can be described by 
a simple power-law with little absorption (\citealt{Margutti2019}), with a photon index in the range $\sim$1.6--1.7, 
indicating no evident spectral evolution. 
Following the method outlined in previous works (\citealt{Dewangan2006, Strohmayer2009}), we used the relation between the 
QPO frequency and power-law index obtained for six systems with dynamic mass constraints
 (\citealt{Vignarca2003, Shaposhnikov2009, Strohmayer2009}) to find the possible range of QPO frequencies with power-law indices 
between 1.6 and 1.7. 
Under the assumption that QPO frequency scales inversely in proportion to BH mass     (\citealt{Remillard2006}), 
we can estimate the black hole mass in \src to be $M_{\rm BH}\sim2.2\times10^{3}$\msun ~(Figure 4). 
The above estimates of the BH mass ($\sim$$10^{3}-10^5$\msun) suggest that \src may harbor an IMBH. 


In order to explain its fast evolving UV and optical emission, (\citealt{Perley2019})
have suggested that \src could be powered by the tidal disruption event (TDE) of a main-sequence star 
by an IMBH. They used the MOSFit TDE model (\citealt{Mockler2019})  to fit the UV and optical light curves 
and obtained a best-fit BH mass of $1.9\times10^{4}$\msun~disrupting a star of 0.6\msun. 
However, it is possible that the early UV and optical emission originates from the circularization process (\citealt{Piran2015}), 
rather than accretion, posing a challenge to apply the MOSFit model which assumes the TDE UV/optical emission 
is related to the reprocessed accretion luminosity. 
We attempted to develop a TDE accreting model to fit its X-ray light curve obtained from Swift-XRT observations. 
In this framework, after a star being disrupted by the black hole, its debris will fall back to pericenter on a time-scale (\citealt{Rees1988})
\begin{equation}
t_{\rm fb} \simeq 4.1\ \left(\frac{M_{\rm BH}}{10^4\ M_\odot}\right)^{1/2} \left(\frac{M_*}{M_\odot}\right)^{-1} \left(\frac{R_*}{R_\odot}\right)^{3/2}\ {\rm day}, 
\label{eq:tfb}
\end{equation}
where $M_{\rm BH}$ is the black hole mass, $M_*$ is the mass of disrupted star and $R_*$ is its radius. 
In the late time, the fallback rate $\dot M_{\rm fb}(t)$ drops with an asymptotic power-law index $-5/3$ 
(\citealt{Ramirez2009, Chen2018}). 
However, the power-law index can approach to $-9/4$ rather than the canonical value of $-5/3$, 
if the stellar core survives after the pericentric encounter (\citealt{Guillochon2013, Coughlin2019}) or 
the disrupted star is in its late age (\citealt{Golightly2019}). 
The luminosity history will trace the mass fallback rate if the fallback mass is rapidly accreted by the black hole. In this case, accretion rate $\dot M_{\rm acc}$ will be close to the fallback rate $\dot M_{\rm fb}$, and the luminosity is $L(t) = \eta \dot M_{\rm fb} c^2$, where $\eta \ll 1$ is the efficiency of  converting accretion power to luminosity.
However, this might not be occurring if the accretion time-scale is longer than the fallback time-scale. 
In this case, the relation between the accretion rate and the mass fallback rate can be given by (\citealt{Chen2018, Kumar2008, Lin2017, Mockler2019})
\begin{equation}
\dot M_{\rm acc} (t) = \frac{1}{\tau_{\rm acc}} \left(\rm e^{-t/\tau_{\rm acc}} \int^t_{t_{\rm fb}} \rm e^{t'/\tau_{\rm acc}} \dot M_{\rm fb}(t')\ dt' \right),
\label{eq:M_acc}
\end{equation}
where $\tau_{\rm acc}$ is the so-called ``slowed'' accretion time-scale.

Here we approximate the overall debris fallback history as follows. The fallback rate remains constant at $\dot M_{\rm peak}$ between $t_{\rm fb}$ and $1.5 t_{\rm fb}$, and after $1.5 t_{\rm fb}$ it starts to decay as $t^{-5/3}$, i.e.,

\begin{equation}
\dot M_{\rm fb} = \begin{cases}
0,&t \lesssim t_{\rm fb} \\
\dot M_{\rm peak},&t_{\rm fb} \lesssim t \lesssim1.5 t_{\rm fb} \\
\dot M_{\rm peak} \left( \frac{t}{1.5 t_{\rm fb}} \right)^{-5/3},&t\gtrsim 1.5t_{\rm fb}
\end{cases}
\label{eq:Mfb}
\end{equation}
For the case of full disruption, $\dot M_{\rm peak} \simeq 0.2 M_* / t_{\rm fb}$. 
The luminosity history is $L(t) = \eta \dot M_{\rm acc} c^2$. If the ``slowed'' accretion time-scale is $\tau_{\rm acc} = 0$, 
it becomes $L(t) = \eta \dot M_{\rm fb}(t) c^2$.
We used the parameter $\eta$, black hole mass $M_{\rm BH}$, and TDE start time $t_0$ to fit the X-ray light curve. 
The disrupted star is assumed to be solar-type.  
The best fitting result yields $M_{\rm BH} \sim 10^4\ M_{\odot}$, $t_0 \sim 58281$ (Modified Julian Date days), $\tau_{\rm acc} \sim 9$ day, and $\eta \sim 10^{-5}$. 
The corresponding fallback time-scale is $t_{\rm fb} \sim 4$ days. 
The best-fit light curve is shown in the Figure 5 for the late-time luminosity evolution parameterized with $t^{-5/3}$ decline law (black), 
and with $t^{-9/4}$ decline law (red), respectively.  

\begin{figure*}[t!]
\centering
\includegraphics[width=\linewidth]{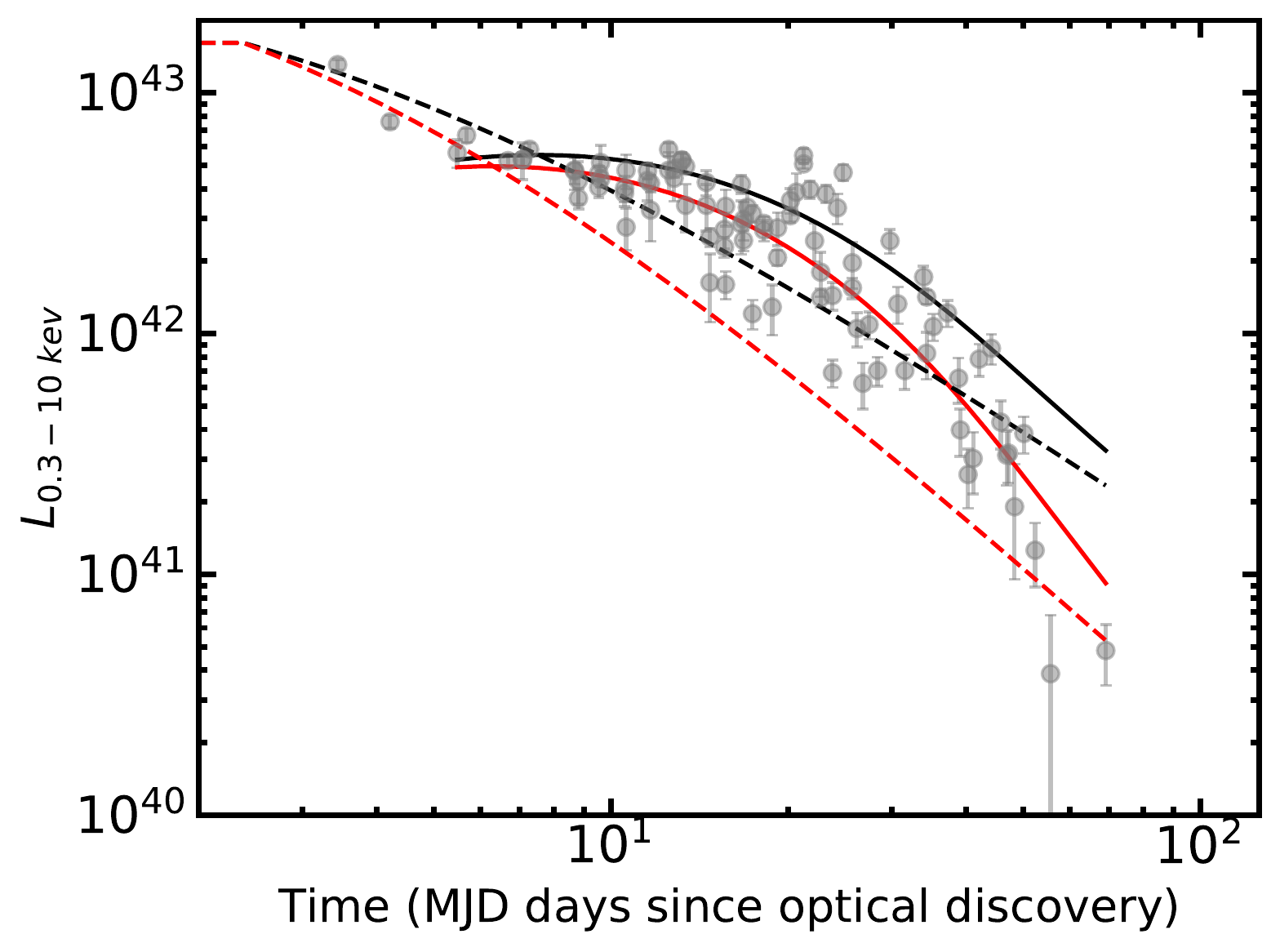}
\caption{The best-fit result to the X-ray light curve  using the TDE model. 
The X-ray luminosity was derived using the data from Swift-XRT observations (Section 2). 
The solid line represents the best-fitting light curve for the viscously slowed accretion model. 
The dashed line corresponds to the light curve with $\tau_{\rm acc} = 0$. 
The black and red lines are for the luminosity evolution models assuming a power-law index of $L\propto t^{-5/3}$ and $L\propto t^{-9/4}$ at later times, respectively. 
The latter steeper power-law index is predicted if the stellar core survives after the pericentric encounter (\citealt{Guillochon2013,Coughlin2019}) or the disrupted star is in its late age (\citealt{Golightly2019}).  
The best-fitting gives the ``slowed'' accretion time-scale $\tau_{\rm acc} \sim 9$ days, the black mass $M_{\rm BH} \sim 10^4\ M_{\odot}$, and the corresponding fallback time-scale is $t_{\rm fb} \simeq 4$ days if the disrupted star is solar-type. 
}
\end{figure*}

The best-fit TDE model yields a BH mass of $\sim10^4$\msun, which is consistent with the mass range constrained by the QPO analogy, 
as well as that obtained by fitting the MOSFiT model to the UV/optical emission of \src (\citealt{Mockler2019}). 
 Given the observed peak bolometric luminosity of $\sim4\times10^{44}$\erg (\citealt{Margutti2019}), such a black hole mass 
 requires that the system to be radiating at about 100 times the Eddington limit.  
 On the other hand, the X-ray emission may be anisotropic and beamed, given the detections of 
 fast-evolved radio radiation (\citealt{Margutti2019, Mohan2020, Bietenholz2020}). 
 The X-ray timing and spectral properties, i.e., QPO frequency, rms amplitude, and power-law dominated X-ray spectrum, 
 make \src similar to the jetted TDE Sw J1644+57 (\citealt{Reis2012}), though the latter has a much higher luminosity, 
 indicating that similar processes may operate in two objects.


We note that \citet{Pasham2021} identified a soft X-ray QPO 
at a much higher frequency of 224 Hz, setting a tight mass limit for compact object in \src to be less than 850\msun~if due to a black hole. 
If the signal were true, the tension can be alleviated by introducing a binary compact object system instead of a single neutron 
star or black hole. 
Such an exotic system comprised of a stellar-mass compact object 
orbiting around an IMBH with mass $\sim10^{3}-10^{5}$\msun~, if confirmed, would be valuable gravitational wave sources 
for future space-based missions such as Laser Interferometer Space Antenna (\citealt{Babak2017}).
In conclusion, our results suggest that the FBOT \src may be powered by an IMBH residing in young 
star clusters off-centre from its dwarf host galaxy (\citealt{Morokuma2019}).  
Although a variety of searching strategies have been proposed (\citealt{Noyola2008, Farrell2009, Lin2018, Greene2020}), 
the existence of IMBHs is still in dispute. 
Luminous FBOTs are promising candidates in future search for IMBHs, 
which will provide further insights into their formation mechanisms. 

\begin{acknowledgements}
This research made use of the HEASARC online data archive services, supported by
NASA/GSFC. We thank the PI who proposed the \xmm and \swift observations of \src  
for making the data available. 
 This work is supported by Chinese NSF through grant Nos. 11822301, 12192220, 12192221, and 11833007.
\end{acknowledgements}




\bibliographystyle{raa}
\bibliography{bibtex}


\clearpage

\appendix

\section{Comparison of XMM-Newton PDS at different energy bands}
\setcounter{figure}{0}
\renewcommand{\thefigure}{A\arabic{figure}}

 Figure A1 shows the XMM/PN PDS extracted in different energy bands. 


 \begin{figure*}[htb]
\centering
\includegraphics[scale=0.3]{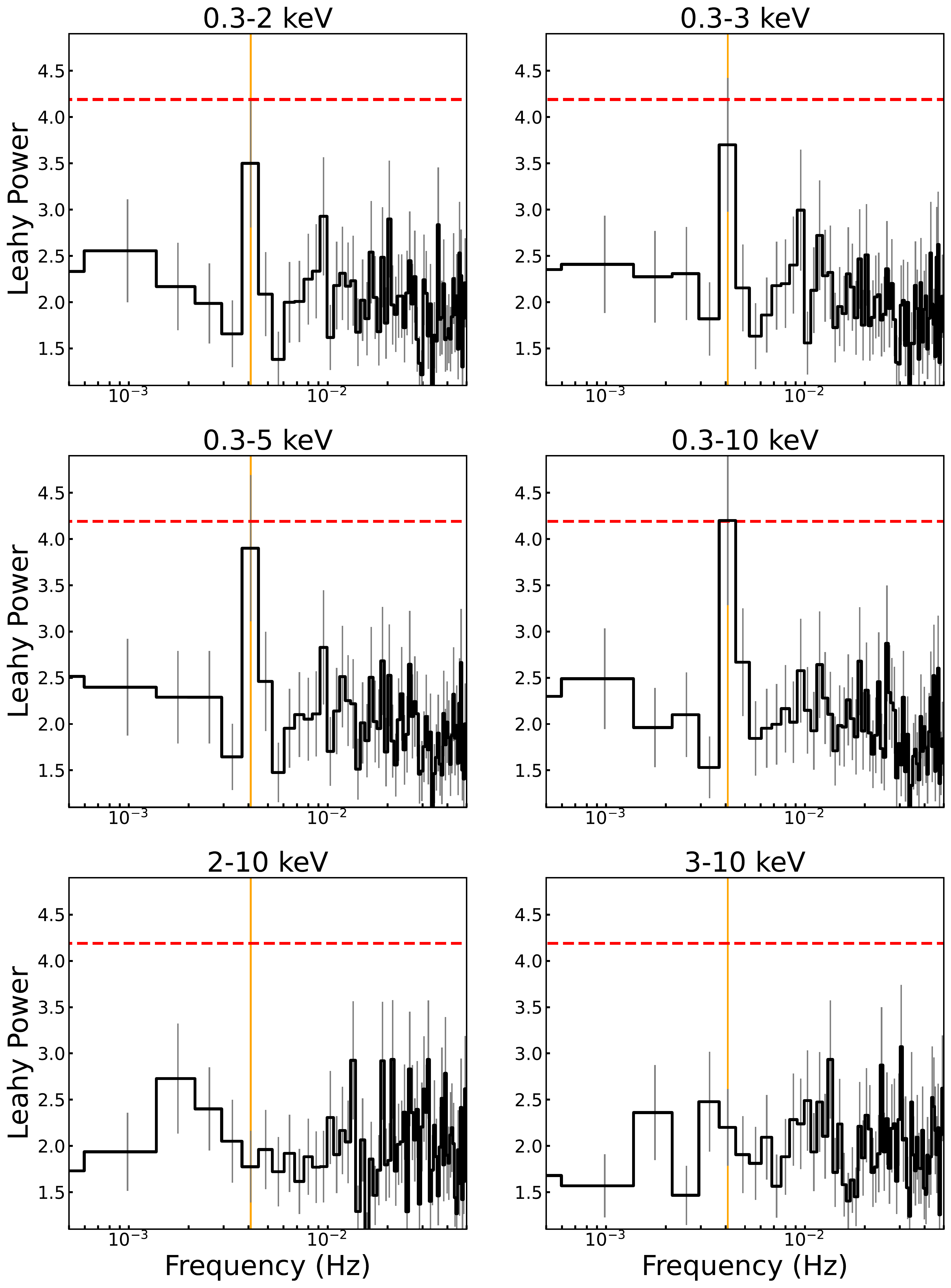}
\caption{
The PDS of XMM-Newton/PN light curve constructed at six different energy bands. 
The red dashed line represents the power value corresponding to the {99.76\%} confidence level,   
while the orange line shows the QPO frequency at 4.1 mHz. 
}
\end{figure*}

\section{Swift-XRT data used in the PDS analysis}
\setcounter{table}{0}   
\renewcommand{\thetable}{B\arabic{table}}

 Table B1 lists \swift data used in the PDS analysis (Section 3). 
For the observation of S11, the light curve can be divided into two segments with with 1.25 ks. 
All other $\geq$2.5 ks observations can not be divided into two segments due to the presence of gap in the light curve. 
Note that we extracted the average PDS from the light curve segments for the NuSTAR and NICER observations, 
but did not detect the QPO signal at the 4.1 mHz.  
This is possibly due to the fact that the QPO signal is dominated by photons at soft bands (Figure A1), 
while only data in the hard X-ray band at $>3$ keV can be used for NuSTAR observations. 
For NICER observations, most light curve segments have too short exposures ($\simlt$1 ks) to 
be used effectively in searching for the QPO signal through stacking PDS. 
Hence, we do not display the NuSTAR and NICER data in this paper.  

\begin{table*}[!b]
  \centering
  \caption{Swift-XRT data used in the PDS analysis}
  \begin{tabular}{c c c c c c}
    \hline 
    \hline 
NO. & ObsID & Exposure (ks) & Obs.Date & Days & Segment    \\
    \hline  
S1 & 00010724009 & 1.3 & 2018-06-22 & 6& 1 \\
S2 & 00010724010 & 1.4 & 2018-06-23 &7& 1 \\
S3 & 00010724019 & 1.3 & 2018-06-25 & 9&1 \\
S4 & 00010724037 & 1.4 & 2018-06-29 &13 &1\\
S5 & 00010724041 & 1.3 & 	2018-06-30 &14 &1 \\
S6 & 00010724043 & 1.3 & 	2018-07-01  &15& 1 \\
S7 & 00010724051 & 1.2 & 2018-07-03 &17 &1 \\
S8 & 00010724054 & 2.1 &2018-07-04 &18& 1 \\
S9 & 00010724057 & 2.7& 2018-07-05  &19& 1 \\
S10 & 00010724060 & 2.5 & 2018-07-06 &20& 1\\
S11 & 00010724063	& 2.8 & 2018-07-07 &21& 2 \\
S12 & 00010724067 & 2.5 & 2018-07-08 & 22&1 \\
\hline
\hline

  \end{tabular}
\end{table*}

\section{Tests for noise properties in the PDS of XMM-Newton and \swift data}
\setcounter{figure}{0}
\renewcommand{\thefigure}{C\arabic{figure}}

 Figure C1 (upper row) shows the cumulative distribution function (left) and probability density function (right) of noise powers in the PDS of XMM-Newton data. 
It is evident that the observed noise powers are consistent with $\chi^{2}$ distribution (with $2\times21$ d.o.f, see Section 3 for details) scaled by a factor of 1/21, i.e., consistent with white noise. 
Note that we performed similar analysis of noise properties in the PDS of \swiftp/XRT data, and found the noise power values are also $\chi^{2}$ distributed (lower row).

 \begin{figure*}[htb]
\centering
\includegraphics[scale=0.6]{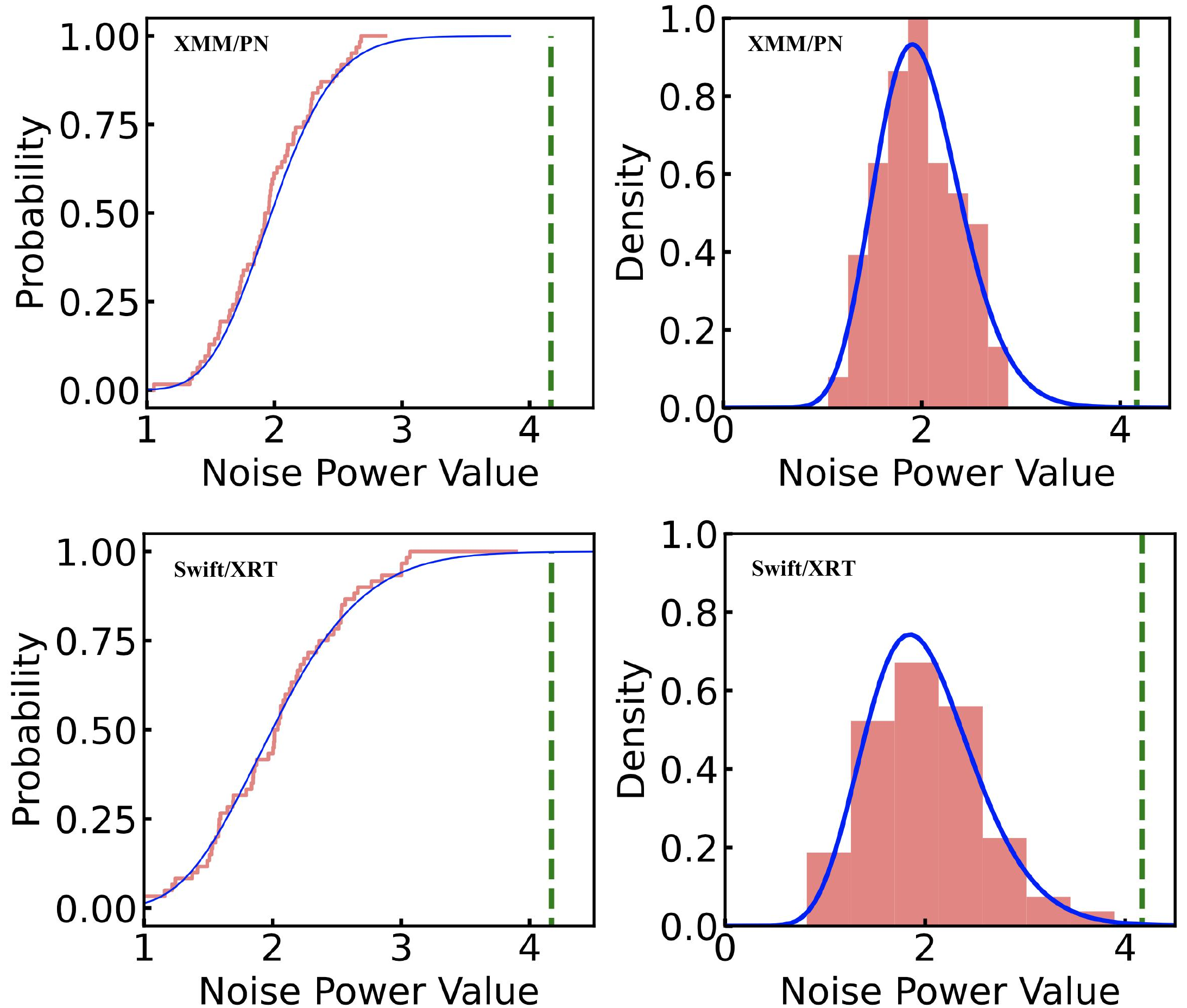}
\caption{Tests for noise properties in the PDS of XMM-Newton (upper row) and \swift (lower row) data.  Left: the red step line shows the cumulative distribution function (CDF) of noise power below 0.05 Hz. 
The three frequency bins centered at 4.1 mHz were removed. The blue curve represents the expected $\chi^{2}$ distribution. 
It can be seen that the blue curve is in good agreement with the CDF of the observational data.  Right: the red histogram shows the density distribution of noise powers, which is consistent with the expected $\chi^{2}$ 
distribution for white noise (blue curve). 
The green dashed line indicates the power value at 4.1 mHz for XMM/PN and Swift/XRT, respectively. 
}
\end{figure*}


\section{Tests on the detectability of the 4.1 \lowercase{m}H\lowercase{z} QPO in the \xmmp/MOS data}
\setcounter{figure}{0}
\renewcommand{\thefigure}{D\arabic{figure}}

We reduced the data from \textit{XMM-Newton}/MOS observations (MOS1 and MOS2) 
following the standard procedures using the \textit{XMM-Newton} SAS v17.0.0, with the latest calibration files.  
Similar to PN, the source light curves in the energy range of 0.3--10 keV were extracted using a circular region with a radius of 35\arcsec, 
while the background light curves were extracted from source-free areas on the same CCD using four identical circular regions. 
No strong background flares were present during the MOS observations.  
Figure D1 (upper panel) shows the MOS1, MOS2 and MOS1+MOS2 light curves. 
Then we produced the corresponding Leahy-normalized PDS with a frequency resolution of 0.8 mHz (Figure D1, lower panel), where 
no strong feature at 4.1 mHz can be seen, with the power values being consistent with 
the noise level of $\sim$2. 
This suggests that even if we combine the PN and MOS data, the signal would not necessarily be getting stronger ({Figure D2}).

\begin{figure*}[htb]
\centering
\includegraphics[scale=0.4]{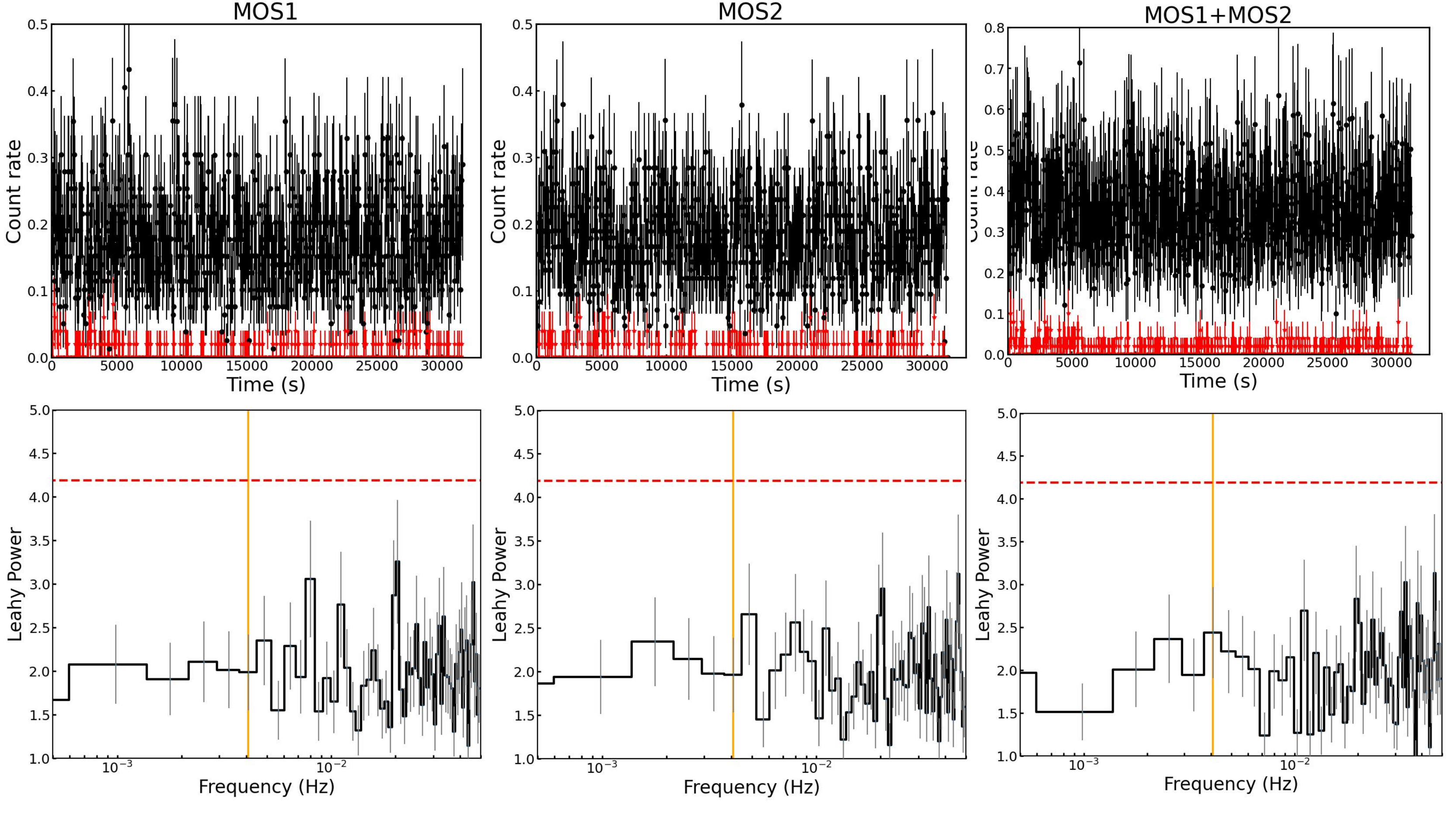}
\caption
{\xmm MOS1, MOS2 and MOS1+MOS2 light curves (upper panel), and the corresponding PSD (lower panel). The background light curves are shown in red. 
The red dashed line represents the 99.76\% confidence level for the QPO detected at the frequency of 4.1 mHz in the PN data (orange vertical line, 
see Figure 1 for details).
}
\end{figure*}

\begin{figure*}[htb]
\centering
\includegraphics[scale=0.6]{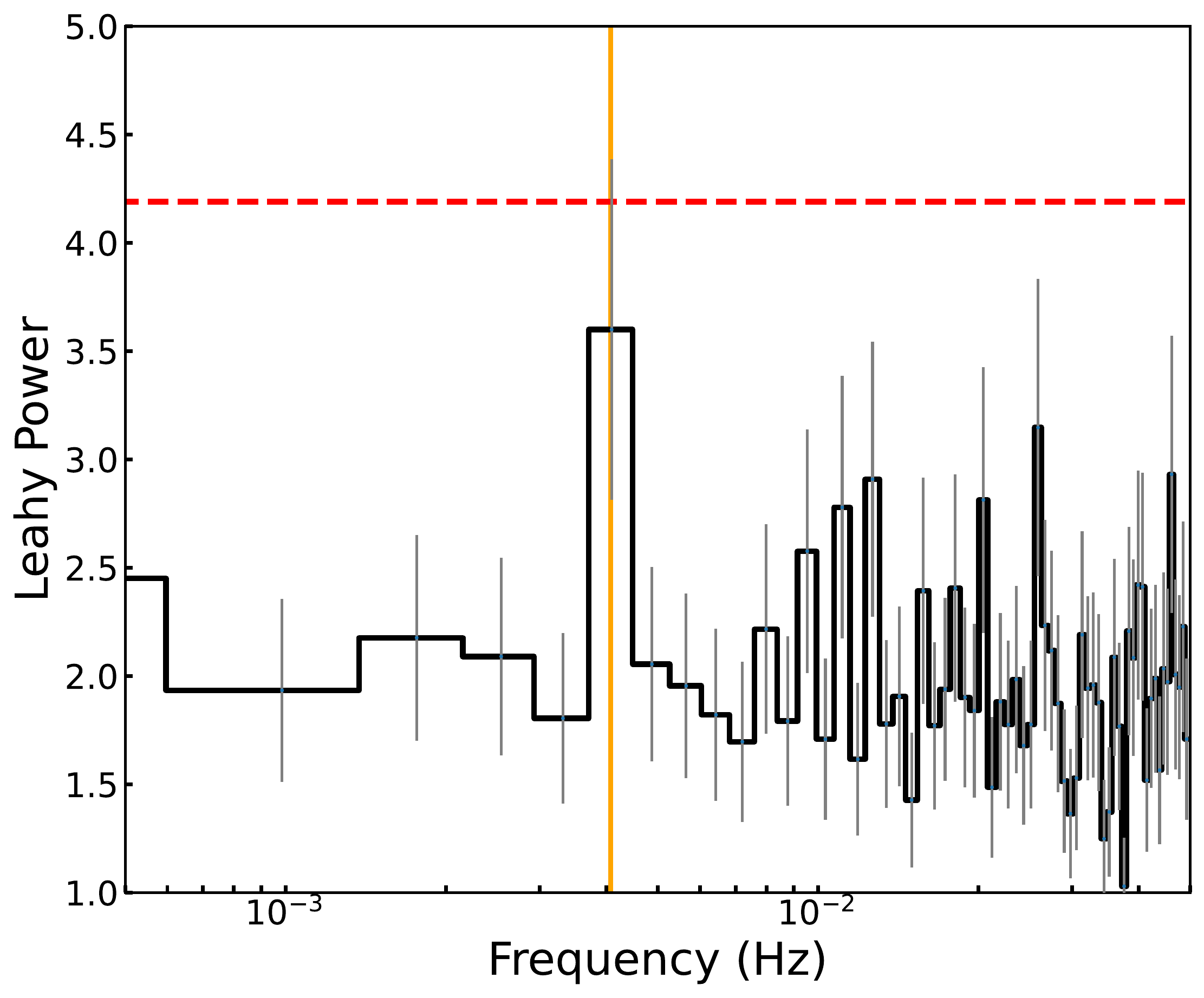}
\caption
{The Leahy-normalized PDS constructed by combining PN and MOS light curves.
}
\end{figure*}

Furthermore, we performed simulations to assess whether the 4.1 mHz QPO signal of AT2018cow seen in the PN data (at {99.76\%}) 
is indeed not detectable by MOS. We generated fake MOS1 light curve as follows. We first divided the observed source light curve of PN by a factor of 3.5 
(the ratio of mean count rates between PN and MOS1), and added the observed background light curve of MOS1 to resemble the expected 
source+background light curve of MOS1. By assuming Poisson distribution, we then re-sampled the expected counts in each time bin. Finally, 
we subtracted the observed background counts from the expected counts to produce fake source light curve of MOS1. 
We also generated fake MOS2 light curve in a similar way. 
We repeated the procedures for $10^{4}$ times and obtained  $10^{4}$ simulated MOS1+MOS2 light curves and PDS. Among  $10^{4}$ PDS, 
only 7 have the power values at 4.1 mHz reaching the {99.76\%} level as detected in PN (Figure D3). 
This suggests that the chance of QPO detection in the MOS data is indeed low ($<$0.1\%). 

\begin{figure*}[htb]
\centering
\includegraphics[scale=0.6]{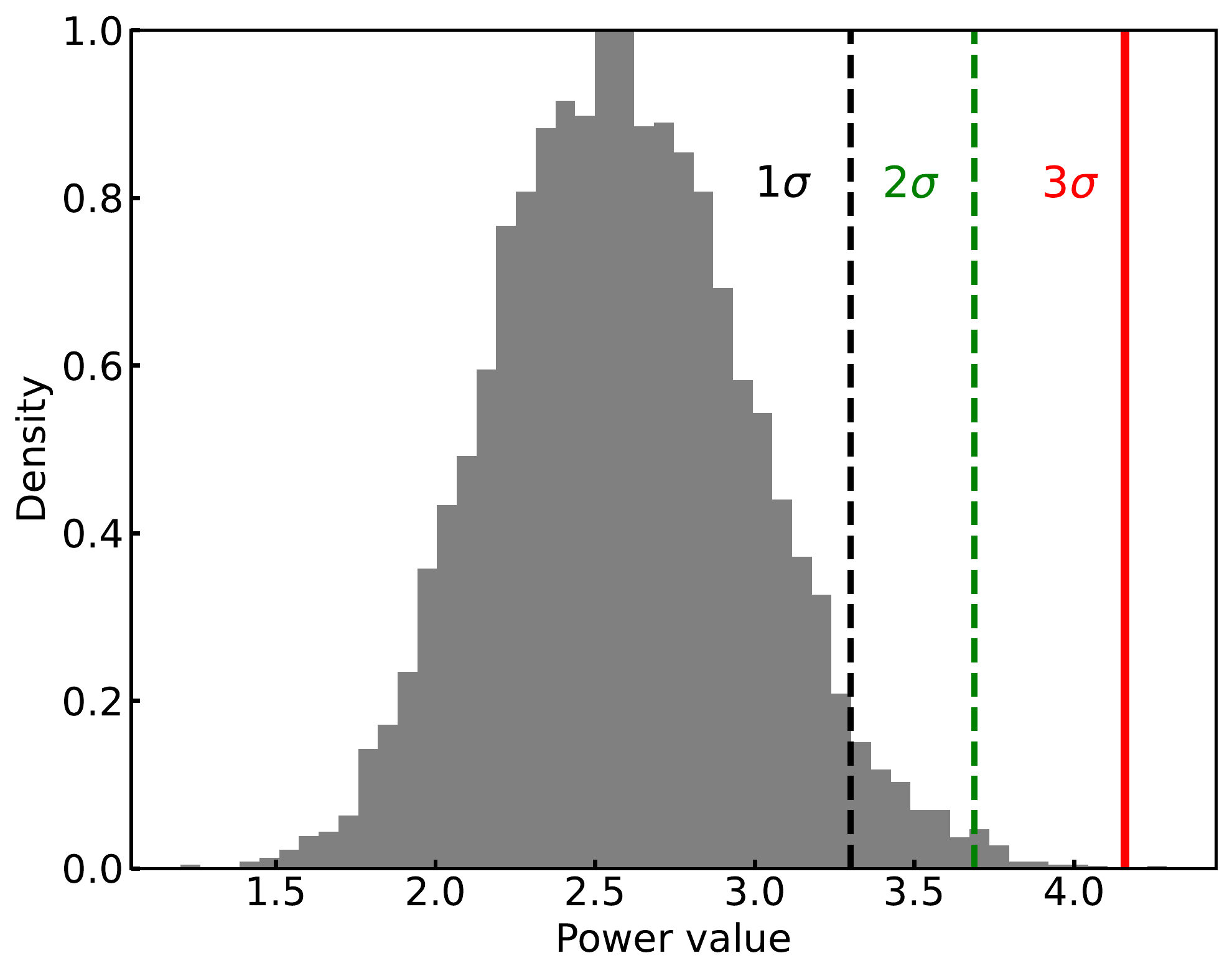}
\caption{
Distribution of power values at 4.1 mHz from light curve simulations by decreasing the PN count rates to mimic the MOS (MOS1+MOS2) ones. The vertical lines represent the 1, 2, 3$\sigma$ confidence levels obtained from the simulated MOS PDS.}
\end{figure*}

\end{document}